\documentclass[twocolumn,showpacs]{revtex4}

\def\D#1#2{\frac{\partial #1}{\partial #2}}
\def\d#1#2{\frac{d\, #1}{d\, #2}}

\def\h{\frac{1}{2}}
\def\s{\sin^2\vartheta}
\def\ctg{\mathrm{ctg}}

\begin{document}
\title{Cosmic Vortexes in Global Time Theory}
\author{D.E. Burlankov}
 \altaffiliation{Physics Department, University of Nizhny Novgorod, Russia.}
 \email{bur@phys.unn.runnet.ru}
%\affiliation{%
%Authors' institution and/or address}

\begin{abstract}
Introduced herein are the foundations of {\it Global Time Theory} (GTT) that further
develop the General
Relativity (GR). GTT is significantly different from GR in the general
physical concepts, but retains 90\% of the mathematical structure and main results.
Meanwhile, description of the cosmos dynamics in GTT leads to significant modifications.
The dynamics equations are derived from Lagrangian, and the Hamiltonian of gravitation is nonzero.
Detailed solutions to the cosmic vortexes, that possess a weak principle of
superposition and a huge energy, are derived. The virial theorem of space
is formulated and proved.
\end{abstract}

\pacs{04, 04.20.Cv}
\maketitle
\tableofcontents

\section{Introduction}\

    Currently, the accepted theory of space, time and gravity is General Relativity (GR). This theory
examines space and time as a four-dimensional repository that can be modified (in accordance
with Einstein's equations) by the matter contained therein. As in the times of Mach,
the basis of cosmic dynamics is ``tangible matter''.

However, astrophysical observations of XX century established that dynamics of
galaxies and heaps of galaxies cannot be explained based solely on gravitational
interactions of visible stars. In order to explain the observed anomalies notions of ``dark matter'', ``dark energy'', massive ``dark holes'',
etc. were imagined. Compared to Newtonian dynamics, GR as the space-time theory
suggests only small corrections (at this scale).

Further development of GR lead to {\it Global Time Theory} (GTT), (see \cite{Bjern}) which significantly
differs  from GR in base physical postulates, but retains 90\% of the mathematical structure and main results.
Meanwhile, description of the cosmos dynamics in GTT leads to significant modifications.

\section{Global Time Theory}\

In GTT time is absolute. It flows equally -- always and everywhere -- and is itself the
measure of equality. The development of {\it the Entire Universe} occurs in this global time.

The space has three dimension, is Riemanian, and its metric tensor ($\gamma_{ij}$)
can depend on space coordinates and time. Points of space are linked with it absolutely. The
frame of reference in which coordinates of space points do not change is called {\it
the global inertial system}.

The inertial system permits arbitrary tree-dimensional transformations of coordinates $\tilde{x}^i(x^j)$
(that do not depend on time).

The coordinate transformations that are {\it  time-dependent}  lead to {\it global non-inertial
system} of observer. Meanwhile the time remain global. In the non-inertial system
the {\it vector field of absolute velocities} $V^i$ arises, although it vanish in the inertial system.
In the transformation of coordinates, the $V^i$-field is transformed as a gauge field:
\begin{equation}
\tilde{x}^i=f^i(x,t);\quad
\tilde{V}^i=\D{\tilde{x}^i}{t}+\D{\tilde{x}^i}{x^k}V^k;\quad
\tilde{\gamma
}^{ij}=\D{\tilde{x}^i}{x^k}\D{\tilde{x}^j}{x^l}\gamma ^{kl}.
\label{pre}
\end{equation}

\subsection{The covariant derivative over the time}\

We will denote the time derivative in the inertial system as  $D_t$ and call it
{\it the covariant derivative over the time}. By the rule of the composite function
differentiation
\begin{equation}
D_t\:F=\D{F}{t}+\D{F}{x^i}\:\D{x^i}{t}=\D{F}{t}+V^i\D{F}{x^i},
\label{covtscal}
\end{equation}
which determines the covariant derivative over the time of scalar field (action, eikonal) in
an arbitrary frame with global time.

The structure of  covariant derivatives over the time of tensors acquires additions
in the form of Lie-variation, that are generated by transformation of coordinates  $\delta x^i=-V^i\,dt$ --
for return  to the inertial system.

For a tensor of an arbitrary range
 \begin{equation}
D_t\:Q^i_{jk}=\D{}{t}Q^i_{jk}-V^i_{;s}Q^s_{jk}+V^s_{;j}Q^i_{sk}+
V^s_{;k}Q^i_{js}+V^sQ^i_{jk;s}. \label{covtens}
\end{equation}

Especially important for the theory is the covariant derivative over the time of the
metric tensor:
\begin{equation}
D_t\:\gamma _{ij}=\D{\gamma _{ij}}{t}+V_{i;j}+V_{j;i}.
\label{difmtr}
\end{equation}

\subsection{Action and dynamical equations}\

In GR, space play a rather passive role.
In GTT the three-dimensional space  is the dynamic field, relative to which there
exists an absolute motion, or, on the contrary, there exists a field of space
velocities in a given system of coordinates. The equations of motion
are derived from the variational principle. Lagrangian is presented as the difference
of the kinetic and potential energy.
Introducing the {\it tensor of space deformation velocity}
\begin{equation}
\mu _{ij}=\frac{1}{2\,c}D_t\,\gamma _{ij}=\frac{1}{2\,c}(\dot{\gamma}_{ij}+
V_{i;j}+V_{j;i}), \label{muij}
\end{equation}
we can represent this action as preconcerted with the Hilbert action in GR:
\begin{equation}
 S=\frac{c^4}{16\,\pi\,k}\int(\mu^i_j\,\mu^j_i-(\mu^j_j)^2+R)
\sqrt{\gamma}\,d_3\,x\,dt+ S_m,
\label{Lagr}
\end{equation}
where $S_m$ is the action of enclosed matter, which adds to dynamic equations the
energy-momenta tensor components. The absolute velocities $V^i$ are represented only
in kinetic energy.

By introducing momenta
$$\pi^i_j=\frac{\sqrt{\gamma}}{2}\,(\mu^i_j-\delta^i_j\,\mu^s_s),$$
and varying the action by six components of spacial metrics, we obtain the six
equations of dynamics:
\begin{equation}\label{qij}
\dot{\pi}^i_j=b^i_j+\sqrt{\gamma}\,G^i_j+\sqrt{\gamma}\,(T^i_j-V^i\,T^0_j),
\end{equation}
where $b^i_j$ we call {\it the self tensor current}
\begin{equation}\label{self}
 b^i_j=-\delta^i_j\,\frac{1}{\sqrt{\gamma}}(2\,\pi^k_l\pi^l_k-\pi^k_k\pi^l_l)-
\partial_s(V^s\,\pi^i_j)+V^i,_s\,\pi^s_j-V^s,_j\,\pi^i_s,
\end{equation}
$G^i_j$ is the Einstein's tensor of three-dimensional space, and $T^\alpha_\beta$ --
components of tensor energy-momenta of enclosed matter, which
determine {\it the exterior tensor current}.

In three-dimensional space, the Riemann-Kristoffel tensor is algebraically expressed
through the Einstein's tensor ($\epsilon_{ijk}$ -- absolute antisymmetric tensor):
$$R^{ij}_{kl}=-\epsilon^{ijs}\epsilon_{klm}G^m_s,$$
and therefore the absence of currents leads to a flat space.

The variation (only the kinetic part of action) by three components of the field of
absolute velocities gives three equations of links:
\begin{equation}\label{varV}
  \nabla_i\,\pi^i_j=\sqrt{\gamma}\,\frac{8\pi k}{c^4}T^0_j.
\end{equation}

These equations are linear along $V^i$ and can be represented by
\begin{equation}\label{links}
%  \frac{1}{2}\nabla^i\partial_t\gamma_{ij}-
% \frac{1}{\sqrt{\gamma}}\,\partial_j\partial_t\sqrt{\gamma}+
  \frac{1}{2}\nabla^i\dot{\gamma}_{ij}-
  \frac{1}{\sqrt{\gamma}}\,\partial_j\dot{\sqrt{\gamma}}+
  R_{ij}\,V^i+\frac{1}{2}\nabla^i(V_i,_j-V_j,_i)=\frac{8\pi k}{c^4}T^0_j.
\end{equation}

The Hamiltonian differs from the Lagrangian only by the sign in front of the potential component of
energy:
\begin{equation}\label{HamGr}
  H=\frac{c^4}{16\pi k}\int\left(\frac{\pi^i_j\pi^j_i-(\pi^i_i)^2/2}{\sqrt{\gamma}}-
  R\,\sqrt{\gamma}\right)\,d_3x.
\end{equation}
Its unique feature is the non fixed sign, as a result of which such phenomena as
Friedman cosmological expansion are possible.

\subsection{The proper time of the moving body}\

Similar to GR, the GTT encompasses special relativity. On level with global time,
in which the
growth of Entire Universe occur, for the moving observer there are his local system
and {\it local time}. All phenomena in the moving system develop in this local time.
This can be expressed  through the square of absolute velocity
\begin{equation}\label{loctime}
d\tau=dt\,\sqrt{1-\gamma_{ij}(\dot{x}^i-V^i)(\dot{x}^j-V^j)}.
\end{equation}

This expression can be represented in four-dimensions by joining the time and
space into unified four-dimensional manifold with metrics
\begin{equation}\label{fourmtr}
  g_{00}=1-\gamma_{ij}\,V^i\,V^j;\quad g_{0i}=\gamma_{ij}\,V^j;\quad
  g_{ij}=-\gamma_{ij}.
\end{equation}
The reverse metric tensor of this four-dimensional manifold is
$$g^{00}=1;\quad g^{0i}=V^i;\quad g^{ij}=V^i\,V^j-\gamma^{ij}.$$

The first equation carries great significance
\begin{equation}\label{Glob}
g^{00}=1.
\end{equation}
This is {\it the main structural relationship} in GTT, analogous to the Minkowski
metric, which is the main structural relationship in special relativity.

\subsection{General Relativity}\

If there is the four-dimensional metric $g_{\alpha\beta}$ in arbitrary four coordinates
$x^\alpha,\,\alpha=0..3$, the variable $\tau$ must be determined for reduction to
global time, in order for the main structure relation (\ref{Glob}) to hold true. We
must transfer the metrics component $g^{00}$ by rule of tensor transformation:
\begin{equation}\label{HJGTT}
  \bar{g}^{00}=g^{\alpha\beta}\D{\tau}{x^\alpha}\D{\tau}{x^\beta}=1.
\end{equation}
This differential equation turns out to be Hamilton-Jacoby differential equation for free-falling bodies
(laboratories), the common time for which is $\tau$, which is the global time.

Thus {\it the equivalence principle} is realized, but in contrast to GR, the time of
the inertial system exists not only for a local laboratory, but for a great many
laboratories in all of space.

For example, Kerr's metric \cite{Chandra} in global time has a radial and an angular component
of the absolute velocity field:
\begin{equation}\label{vfi}
  V^\varphi=-\frac{2\,a\,M\,r}{w};\quad V^r=\frac{\sqrt{2\,M\,r(r^2+a^2)}}{\rho^2},
 \end{equation}
where
$$\rho^2=r^2+a^2\,\cos^2\vartheta;\quad w=(r^2+a^2)\rho^2+2\,M\,r\,a^2\,\sin^2\vartheta,$$

The space metrics
\begin{equation}\label{KerrDiag}
\gamma_{11}=\frac{(\rho^2)^2}{w};\quad \gamma_{22}=\rho^2;\quad
\gamma_{33}=\frac{w}{\rho^2}\,\sin^2\vartheta
\end{equation}
has singularity only at $\rho^2=0$.

\subsection{The energy-momentum tensor of the space}\

GTT does not mathematically coincide with GR only in one equation:
since the main structure relation (\ref{Glob})  $g^{00}=1$ not permit to vary this
component, the tenth Einstein's equation, determined by this component variation
in GTT is absent. As a result, the difference
\begin{equation}
-2\:\frac{\delta\:S}{\delta\:g^{00}}=\frac{c^4}{8\pi k}\,G_{00}-T_{00}\equiv\rho
;\quad \label{ten}
\end{equation}
is nonzero. In tensor terms we will denote this difference as {the difference tensor}:
\begin{equation}
\theta ^\alpha _\beta =\frac{c^4}{8\pi k}\,G^\alpha _\beta -T^\alpha
_\beta, \label{dif}
\end{equation}
and as  consequence of (\ref{ten})
 \begin{equation}\label{dust}
\theta ^0_0=\rho ;\quad \theta ^i_0=\rho V^i.
\end{equation}

Since  $G^\alpha_\beta$ and $T^\alpha_\beta$ are subject of
Hilbert's identities, the difference tensor $\theta^\alpha_\beta$
is also subject to them:
\begin{equation}
\nabla_\alpha\theta^\alpha_\beta=0,
\label{divtens}
\end{equation}
thus (\ref{dust}) has the form of the energy-momenta tensor of dust matter.
But if we want to model GTT  in GR by means of dust, as a result of a non fixed
sign of energy density, the possibility of a negative  density of dust in GR
must also be considered.

\section{Solutions}\

The most essential difference between GTT and GR is the nonzero Hamiltonian. The dynamic
equations conserve the Hamiltonian density, and due to its non fixed sign,
partial solutions with all-around density of zero are possible.
These are the GR - solutions, which comprise {\it a subset} of GTT solutions.

We demonstrate a small set of GTT-solutions, which illustrate the solving methods
as well as the similarities and contrasts with GR.

\subsection{The Spherical Universe dynamics}\label{Friedman}\

The simplest model is a three-dimensional sphere with time-depending radius:
$$dl^2=r^2(t)\,ds_3^2,$$
where $ds_3^2$ -- is the metric of the three-dimensional sphere with a radius of one.

For three-dimensional sphere with radius $r$, the scalar curvature
$$R=\frac{3}{r^2},\quad \sqrt{\gamma}=r^3.$$

The kinetic energy is proportional to
$$T=-3\left(\frac{\dot{r}}{r}\right)^2\,r^3=-3\,r\,\dot{r}^2$$
and Hamiltonian is negative:
\begin{equation}\label{FridHam}
    H=-3\,r\,(\dot{r}^2+1).
\end{equation}

The Hamiltonian conservation leads to a differential equation of the first order:
$$-H=3\,r\,(\dot{r}^2+1)=3\,r_{max},$$
which is the Friedman's equation, that have a cicloida solution. In contrast to the
classical formulation of Friedman's problem, this solutions is a vacuum one, without matter,
and $r_{max}$ is the integration constant, which not depended on some matter density.

\subsection{The field of spherical mass}\

The inertial system is dynamical, but in global time there exist solutions, that are
static from the point of view of some noninertial system.

In a spherically symmetric case, the space metric can be transformed to
$$dl^2=dr^2+R^2(r)(d\vartheta^2+\sin^2\vartheta\,d\varphi^2),$$
and the field of absolute velocities is radial: $V^r=V(r)$.
Only the radial link-equation is nontrivial:
$$\nabla_i\pi^i_r=\frac{2}{r}\,R''=0,$$
from which $R=r$, and the space turns out to the flat.

In dynamical equations $q^1_1,\,q^2_2=q^3_3$ are non trivial,
but if the first equations is executed, the second one is completed automatically
as a result of Hilbert identities.
\begin{equation}\label{speq}
  q^1_1=\frac{V\,(2\,r\,V'+V)}{r^2}=\frac{(r\,V^2)'}{r^2}.
\end{equation}

In vacuum $q^1_1=0$, from which
$$r\,V^2=const\equiv 2\,k\,M\geq 0,$$
where $k$ is the gravitational constant, and
$M$ is the constant of integration, which can be treated as the mass of a central body.
This constant must be positive, whereas in GR the positive sign
of mass is a problem.

The field of radial velocities
$$V^r=V=\sqrt{\frac{2\,k\,M}{r}}$$
leads to the four-dimensional metric
$$  ds^2=\left(1-\frac{2kM}{r\,c^2}\right)\,c^2\,dt^2+
  2\sqrt{\frac{2kM}{r}}\,dt\,dr$$
\begin{equation}\label{ShGlob}
- dr^2- r^2(d\vartheta^2+\sin^2\vartheta\:d\varphi^2).
\end{equation}

In 1921 this metric was obtained by Painlev\'e \cite{Painleve} by transformation
of the time variable in Schwrzschild \cite{Chandra} solution of GR.
Painlev\'e's attention  was
attracted by the simplicity of the space section  $t=const$, which turned out to be a
flat Euclidean space.

In the reverse metric of this space the component $g^{00}$ equals one.

\subsection{The Vortex field}\

The task of space vortexes has no analog in GR and is the proper task of GTT.

The metric is stationary, axially-symmetric, and can be transformed into
\begin{equation}\label{rotmtr}
dl^2=e^{w(r,\vartheta)}\,(dr^2+r^2\,d\vartheta^2)+r^2\,\sin^2\vartheta\;d\varphi^2.
\end{equation}

The absolute velocities field is also dependent on $r$ and $\vartheta$, and is
the vortex field $V^\varphi=\Omega(r,z)$.
The space-deformation tensor
$$\mu^3_1=\h \Omega,_r\;\quad\mu^1_3=\frac{r^2}{2}\,e^{-w}\,\Omega,_r\,\s;$$
$$\mu^3_2=\h \Omega,_\vartheta\;\quad
\mu^2_3=\h\,e^{-w}\,\Omega,_\vartheta\,\s;$$
determines the momenta
$$\pi^1_3=\frac{r^4}{2}\,\Omega,_r\sin^3\vartheta;\quad
\pi^3_1=\frac{r^2}{2}\,e^w\,\Omega,_r\,\sin\vartheta;$$
$$\pi^2_3=\frac{r^2}{2}\,\Omega,_\vartheta\sin^3\vartheta;\quad
\pi^3_2=\frac{r^2}{2}\,e^w\,\Omega,_\vartheta\,\sin\vartheta.$$

The kinetic energy
\begin{equation}\label{Kinet}
T=\frac{c^2}{32\pi k}\,\int\left(\Omega,^2_r+\frac{1}{r^2}\,\Omega,^2_\vartheta\right)
\,r^4\,\sin^3\vartheta\,dr\,d\vartheta\,d\varphi
\end{equation}
is determined solely by the vortex field $\Omega$
and is independent of the metric function $w$.

The unique nontrivial link  for $V^\varphi$ in the absence of current
yields the equation for $V^\varphi$:
\begin{equation}\label{eqV}
\Omega,_{rr}+\frac{4}{r}\,\Omega,_r+\frac{1}{r^2}\,\left(\Omega,_{\vartheta\vartheta}+
3\,\ctg\vartheta\,\Omega,_\vartheta\right)=0.
\end{equation}

Note, that this second order linear differential equation is independent
on the metric function $w(r,\vartheta)$.

The proper currents
$$b^1_1=-b^2_2=\frac{1}{4}\,e^{-w}\,\s\,\left(r^2\,\Omega,^2_r-\Omega,^2_\vartheta\right);$$
$$b^1_2=r^2\,b^2_1=\frac{1}{2}\,e^{-w}r^2\s\,\Omega,_r\,\Omega,_\vartheta;$$
$$b^3_3=-\frac{3}{4}\,e^{-w}\,\s\,\left(r^2\,\Omega,^2_r+\Omega,^2_\vartheta\right)$$
together with Ricci tensor, which is determined by metric (\ref{rotmtr})
$$R_{11}=-\frac{1}{2\,r^2}\left(r^2\,w,_{rr}+w,_{\vartheta\vartheta}+
\ctg\vartheta\,w,_\vartheta\right);$$
$$R_{12}=\frac{1}{2\,r}\left(w,_\vartheta+\ctg\vartheta\,r\,w,_r\right)$$
$$R_{22}=\frac{1}{2}\,\left(\ctg\vartheta\,w,_\vartheta-w,_{\vartheta\vartheta}-
2\,r\,w,_r\right),$$
yields equations for metric, from which one can determine the derivatives of
function $w$:
$$w,_r=\frac{r}{2}\left(\Omega,^2_\vartheta-\,r^2\,\Omega,^2_r-
2\,\ctg\vartheta\,r\,\Omega,_r\Omega,_\vartheta\right)\,\sin^4\vartheta;$$
\begin{equation}\label{solw}
w,_\vartheta=\frac{r^2}{2}\left(\ctg\vartheta\left(r^2\,\Omega,^2_r-\Omega,^2_\vartheta
\right)-2r\,\Omega,_r\Omega,_\vartheta\right)\,\sin^4\vartheta.
\end{equation}
Upon completing these relationships along with (\ref{eqV}) for $\Omega$
all equations of dynamics and links are satisfied.

The energy density now is expressed solely trough derivatives of $\Omega$:
\begin{equation}\label{densE}
  \varepsilon\sqrt{\gamma}=\frac{r^2\,c^2}{8\pi k}\,(r^2\,\Omega,^2_r+\Omega,^2_\vartheta)\,\sin^3\vartheta,
\end{equation}
and the kinetic energy is exactly four times smaller. This is the result of {\it the space
virial theorem}.

The full energy in a given region  $B$ without external sources
\begin{equation}\label{EB}
  E_B=2\pi\,\int_B \varepsilon\sqrt{\gamma}\,dr\,d\vartheta
\end{equation}
is positive and reaches a minima in the equation (\ref{eqV}) solutions.

\subsection{The space virial theorem}\

Denoting $\pi^i_i\equiv\pi$, the sum of equations (\ref{qij}), in absence of external
sources (for proper gravitation), gives
\begin{equation}\label{dotspur}
\dot{\pi}+\partial_s(V^s\,\pi)=-3\,T-\frac{1}{2}R\,\sqrt{\gamma}=-3\,T+\,U,
\end{equation}
where $T$ and $U$ are the densities of the kinetic and the potential energy, respectively.
The space virial theorem can be applied to the almost stationery fields in space,
on the boundaries of which there is no current flow, $V^n=0$.
Averaging (\ref{dotspur}) over time, we obtain the relationship between the average
potential, kinetic, and full energies:
\begin{equation}\label{virial}
  U=3\,T;\quad E=T+U=4\,T.
\end{equation}

Under the aforementioned conditions, the kinetic and full energies are positive.

All conditions for application of this theorem to the given task are completed.

For determination of the kinetic energy, it is necessary to know the field of absolute
velocities in a given region. This information may be obtained from visible stars. The
virial space theorem allows calculation of the full energy.

\subsection{The weak superposition principle}\

The main part of the vortex task is to solve the linear differential equation
(\ref{eqV}). Afterwards, equations (\ref{solw}) determine the metric function
$w(r,\vartheta)$.

Although the overall task is nonlinear, the first (main) part -- determination of
the vortex field $\Omega(r,\vartheta)$ -- is linear and subject to the
superposition principle.

Thus, any field $\Omega$ can be represented as the superposition of some basic
solutions. However, equations (\ref{solw}) for finding field $w(r,\vartheta)$ contains
the square of the field $\Omega$ derivatives. The solution as a whole is not a superposition
of partial solutions.

\subsection{Multipole solutions}\

The differential equation (\ref{eqV}) is homogeneous along radius $r$, thus its common solutions can be found in the form of a power series
\begin{equation}\label{mulser}
  \Omega(r,\vartheta)=\sum_{l=0}^\infty \left(A_l\,r^l+\frac{B_l}{r^{l+3}}\right)\,P_l(\cos\vartheta).
\end{equation}
The angular part is subject to the differential equation (where $x=\cos\vartheta$):
\begin{equation}\label{eqPl}
(x^2-1)P_l''+4\,x\,P_l'-l\,(l+3)P_l=0.
\end{equation}

The solutions with whole $l$ are the Hegenbauer's polynomials with $\alpha=3/2$.
They are the base of spherical functions in a five-dimensional space. Particularly,
at $l=-3$ (as at $l=0$) the solution of equation (\ref{eqPl}) is a constant --
there is a {\it monopole solution}
\begin{equation}\label{mono}
  \Omega_0(r,\vartheta)=\frac{1}{r^3}.
\end{equation}

This solution determines the {\it product function} of  polynomials
\begin{equation}\label{produce}
  \frac{1}{(1-2\,s\,x+s^2)^{3/2}}=\sum_{l=0}^\infty\frac{(l+1)(l+2)}{2}\,P_l(x)\,s^l.
\end{equation}
Thus defined, $P_l(1)=1$, as for Legendre's polynomials.

Although these polynomials are well known, we present the first four
$$P_0=1;\quad P_1=x;\quad P_2=\frac{1}{4}(5\,x^2-1);$$
$$P_3=\frac{1}{4}\,x\,(7\,x^2-3);\quad P_4=\frac{1}{8}(21\,x^4-14\,x^2+1).$$
They are orthogonal with weight $(1-x^2)$, and
\begin{equation}\label{norm}
\int_{-1}^1P_l^2(x)\,(1-x^2)\,dx=\frac{8}{(l+1)(l+2)(2\,l+3)}.
\end{equation}

At $x=0$, the product function (\ref{produce}) is
$$
\frac{1}{(1+s^2)^{3/2}}=\sum_{m=0}^\infty(-1)^m\frac{(2m+1)!}{2^{2m}(m!)^2}\,s^{2m}=$$
$$\sum_{l=0}^\infty\frac{(l+1)(l+2)}{2}\,P_l(0)\,s^l,$$
which determines the values of even polynomials at $x=0$ (the odd polynomials are
equal to zero):
\begin{equation}\label{even}
P_{2m}(0)=(-1)^m\frac{(2m)!}{(m!)^2(m+1)2^{2m}}.
\end{equation}

For each {\it separate} multipole solution $\Omega_l=a\,r^l\,P_l(\cos\vartheta)$
 exists a corresponding multipole solution for $w_l$: $w_l(r,\vartheta)=$
\begin{equation}\label{monow}
  a^2\,r^{2(l+1)}\,\sin^4\vartheta\,\frac{1}{4(l+1)}
  (2\,l\,\ctg\vartheta\,P_l\,P_l,_\vartheta-P_l,_\vartheta^2+
  l^2\,P_l^2)
\end{equation}
and an energy density
\begin{equation}\label{monoE}
\varepsilon\sqrt{\gamma}=a^2\,r^{2(l+1)}\,\sin^3\vartheta(P_l,^2_\vartheta+l^2\,P_l^2).
\end{equation}

However, if the vortex field $\Omega$ is the superposition of multipoles, interference
arises and the function $w$ is not the superposition of corresponding monopoles.

\subsection{The solution with ring singularity}\

The task of finding the vortex field in the mathematical part coincides with the task of
the theory of potential in five-dimensional space.
We can immediately say that a sourceless vortex field that is regular and finite
everywhere does not exist.

A natural source for a vortex field is a circular current -- bodies that are rotating
around a common axis similar to the rings of Saturn.
An analogous task in five-dimensional electrostatics is the field of a charged thin
ring (thread) with radius $R$. The task is axis-symmetric, and the vortex field
may be represented by the superposition of multipoles.
Multipole coefficients can be found by expansion of the potential along the axis
(at $\vartheta=0,\,P_l=1$ for all $l$):
$$\Omega(z)=\frac{Q}{(R^2+z^2)^{3/2}}=\left\{\begin{array}{ll}
\frac{Q}{R^3}\sum_{m=0}^\infty A_m\,\left(\frac{z}{R}\right)^{2m}&z<R\\
\frac{Q}{z^3}\sum_{m=0}^\infty A_m\,\left(\frac{R}{z}\right)^{2m}&z>R\\
\end{array}
\right\},$$ where
$$\quad A_m=(-1)^m\frac{(2m+1)!}{2^{2m}(m!)^2}.$$

Since on the axis $z=r,\,\cos{\vartheta}=1$, the vortex field is the
expansion for even multipoles:
$$\Omega=\sum_{m=0}^\infty(-1)^m\frac{(2m+1)!}{2^{2m}(m!)^2}\left(\frac{r}{R}\right)^{2m}\,P_{2m}(\cos\vartheta)$$
at $r<R$ and analogous to outside region. The ring source is in the ``plane'' of
symmetry at $\cos\vartheta=0$, and only the even polynomials effect the expansion.
Substituting in values from (\ref{even}), at $r<R$ we obtain
$$\Omega(r)=\frac{Q}{R^3}\sum \frac{(2m!)(2m+1)!}{(m+1)(m!)^4\,
2^{4m}}\left(\frac{r}{R}\right) ^{2m}.$$ The field and the radius increase
concurrently.

Similarly, at $r>R$
$$\Omega(r)=\frac{Q}{r^3}\sum \frac{(2m!)(2m+1)!}{(m+1)(m!)^4\,
2^{4m}}\left(\frac{R}{r}\right) ^{2m}$$
the field decreases as radius increases.

\subsection{Sources}\

In cosmology, dust-like matter presents a special interest. The tensor energy-momenta
of this matter is
$$T^\alpha_\beta=\rho\,u^\alpha\,u_\beta,$$
where $u^\alpha$ -- 4-vector of velocity.

All components of the vector contain the normalizing factor
$$\lambda=\frac{1}{\sqrt{1-\frac{\gamma_{ij}}{c^2}(\dot{x}^i-V^i)(\dot{x}^j-V^j)}}.$$

Contravariant components of this vector are
$$u^0=\lambda;\quad u^i=\lambda\,\frac{\dot{x}^i}{c}.$$

Covariant components
$$u_0=\lambda\,(1+\frac{\gamma_{ij}}{c^2}(\dot{x}^i-V^i)\,V^j);\quad
u_i=-\lambda\,\frac{\gamma_{ij}}{c}(\dot{x}^i-V^i).$$
Thus
$$u^0\,u_0+u^i\,u_i=1.$$

The vector source in link equations
\begin{equation}\label{T0i}
 J_i= T^0_i=\rho\,u^0\,u_i=-\rho\,c\,\frac{\gamma_{ij}(\dot{x}^j-V^j)}
  {c^2-\gamma_{kj}(\dot{x}^k-V^k)(\dot{x}^j-V^j)}
\end{equation}
and tensor source in dynamic equations
\begin{equation}\label{Tij}
  J^i_j=T^i_j-V^i\,T^0_j=-\rho\frac{(\dot{x}^i-V^i)\,\gamma_{jk}(\dot{x}^k-V^k)}
  {c^2-\gamma_{kj}(\dot{x}^k-V^k)(\dot{x}^j-V^j)}
\end{equation}
are determined by absolute velocity $\dot{x}^i-V^i$ of a moving particle.

Particles at rest relative to space (coherent) do not contribute to sources. For a
particle to function as a source, it must not be coherent, i.e. have a nonzero
absolute velocity.

The dynamics equations of dust-like matter are concerned with free motion of one particle
and mass conservation along the particle trajectory. The particle
motion in a given field of velocities $V^i$ with metric $\gamma^{ij}$ is determined by
 the relativistic Hamilton-Jacoby equation:
\begin{equation}\label{HJ}
  \left(\D{s}{t}+V^i\D{s}{x^i}\right)^2-c^2\,\gamma^{ij}\D{s}{x^i}\D{s}{x^j}=1
\end{equation}
or corresponding Hamilton equations with a Hamiltonian
\begin{equation}\label{Hamiltonian}
  H=\frac{c^2}{2}\gamma^{ij}(x,t)\,p_i\,p_j+\frac{1}{2}\left(1-
 \left(\varepsilon-V^i(x,t)\,p_i\right)^2\right),
\end{equation}
where
$$\varepsilon=-\D{s}{t};\quad p_i=\D{s}{x^i}.$$

In the Hamilton equations
$$  \d{t}{\tau}=-\D{H}{\varepsilon}=\varepsilon-V^i(x,t)\,p_i;$$
\begin{equation}\label{eqHam}
\d{x^i}{\tau}=\D{H}{p_i}=c^2\,\gamma^{ij}\,p_j+V^i\,(\varepsilon-V^i(x,t)\,p_i);
\end{equation}
$$\d{\varepsilon}{\tau}=\D{H}{t};\quad\d{p_i}{\tau}=-\D{H}{x_i}$$
parameter $\tau$ is the proper time of a moving body.
These equations conserve the zero value of the Hamiltonian (energy).

Here, will present only two important solutions without details:
\begin{enumerate}
  \item Coherent particles ($p_i=0;\;\varepsilon=1;\;\dot{x}^i=V^i$)
  at rest relative to space. This solutions is a mathematical expression of the
  {\it first law of Newton}: a particle at rest relative to space maintains its
  state of rest (in common sense,  the concept of constant motion in a curved space
  is absent).
    \item Particles rotating at $\vartheta=\pi/2$
  with absolute (relative to space) angular velocity
   $$\omega_{abs}=\dot{\varphi}-\Omega= r\,\D{\Omega}{r}|_{\vartheta=0}.$$
  Namely these particles are the source of the vortex field in the given task.
 \end{enumerate}

\subsection{From GR view point}\

The nonzero components of Einstein's four-dimensional tensor for multipole solutions
described above are
$$G^0_0=-\varepsilon;\quad G^3_0=-\varepsilon\,\Omega.$$
In GR, such space-time is determined by dust matter with negative density.

For example, at $l=-4$ (the dipole vortex):
$$\Omega=\frac{a\,\cos\vartheta}{r^4};\quad
w=\frac{a^2}{12\,r^6}\,(25\,\cos^2\vartheta-1)\,\sin^2\vartheta.$$
$$\varepsilon=-a^2\,e^{-w}\,\sin^2\vartheta\frac{1+15\,\cos^2\vartheta}{r^8}.$$

\subsection{The Energy}\

To get an idea about cosmic energies, we examine the following task. A globe with
radius $R$ is in constant rotation with angular velocity of $\Omega$ {\it coherently}.
This means that the globe velocity on the surface coincides with the velocities of
space, i.e. the field of angular velocities outside the globe are determined by a
monopole solution
\begin{equation}\label{omsp}
  \omega(r)=\Omega\,\frac{R^3}{r^3}.
\end{equation}
The energy density outside the globe (inside the globe, the field is homogeneous and
the energy density is zero):
$$\varepsilon=\frac{9\,\Omega^2\,R^6\,\sin^2\theta}{r^6},$$
and the full energy of space is:
$$E=\frac{c^4}{16\pi k}\,9\,\Omega^2\,R^6\,2\pi\int_0^\pi\sin^3\vartheta\,d\vartheta
\int_R^\infty\frac{r^2\,dr}{r^6}=$$
\begin{equation}\label{Esphera}
\frac{R^3\,\Omega^2\,c^2}{2\,k}\equiv M\,c^2,
\end{equation}
where $M$ is the equivalent mass (not the mass of the globe)
\begin{equation}\label{massa}
M=\frac{R^3\,\Omega^2}{2\,k}.
\end{equation}

For example, we examine a globe with diameter 20 см. ($R=0.1\,m$), that completes one
rotation per second ($\Omega=2\,\pi\,c^{-1}$). We obtain $M=300\,000\,000$ kg.
To force the space outside the globe to rotate coherently with the globe requires
as much energy as is released upon annihilation of 300 000 tons of matter. Hence,
laboratory experiments with space vortexes are not realistic.

This example also explains why our space is Euclidean with high accuracy: in the
energy expression, there is a huge factor $c^4/(16\,\pi\,k)$ in front of the space
curvature. This means that the smallest deviation from Euclidean space require
tremendous energy.

Our space is (almost) Euclidean not due to the beauty and elegance of Euclidean
geometry, but because this space has minimum energy.

\section{Conclusion}\

The main philosophical difference of GTT from GR is in its rejection of the {\it
principle of general covariancy}. This difference is not in its mathematical treatment
(upon coordinate transformation all equations must be transformed correctly),
but in its physical vulgarization, i.e the confirmation of the principal
{\it physical} equality of all nonsingular coordinate systems, and as a result --
the confirmation of principal nonpossibility of the
introduction of global time. GTT revives the global time, which simplifies and
enriches the theory.

The main {\it physical} difference of GTT from GT is in nonzero energy density. In
contrast to GR, in GTT
\begin{itemize}
  \item the quantum theory of gravitation is built on the basis of the Schroedinger
  equation, as for other fields;
  \item naturally, as for other fields, the Coshi task is formulated;
  \item naturally, sequential approximation method is constructed.
\end{itemize}

If only a subset of solutions with zero energy density is chosen in GTT
\begin{equation}\label{hzero}
  H=0,
\end{equation}
where $H$ is the common density of the Hamiltonian of space and enclosed matter, we
obtain GR, at which (\ref{hzero}) is Einstein's tenth equation.

Therefore, one of the ways to General Relativity is as follows:
\begin{quote}
 {\it  to accept the global time theory, and then request a zero energy density
 from some assumptions}.
\end{quote}

Are there some logical, or better yet, experimental foundations for $H=0$?

In this link, it is interesting to compare once more the obtained series of multipole
vortex solutions which have a (huge!) positive energy in GTT with Kerr's solution and
a series of solutions by Tomimatsu and Sato \cite{Sato} in GR, which have a zero
energy density. In Kerr's solution (\ref{vfi}), the vortex field at infinity decreases
as $1/r^3$, same as in the monopole solution of GTT, which behaves in this manner
everywhere. However, in Kerr's solution, there is a radial component of absolute
velocity which adds a negative energy. The virial theorem, which leads to positive
energy density, is no longer applicable. Very complex solution with zero energy
density are obtained as a result of a rather difficult combination
of two components of absolute velocity and consideration of corresponding changes of
metric components. These are the GR solutions. In contrast, multipole solutions of GTT
are obtained from a {\it linear} differential equation.

Author thanks Ksenia P. Brazhnik for translation.

\end{document}